\documentstyle[11pt,epsfig]{article}
\title{\bf Stabilization of internal space in\\
noncommutative multidimensional cosmology}
\author{N. Khosravi$^1$\thanks{email:
n-khosravi@sbu.ac.ir}, S. Jalalzadeh$^1$\thanks{email:
s-jalalzadeh@sbu.ac.ir}
  and H. R. Sepangi$^{1,2}$\thanks{email:
hr-sepangi@sbu.ac.ir}
\\ $^1${\small Department of Physics, Shahid Beheshti University, Evin,
Tehran 19839, Iran}\\$^2${\small Institute for Studies in
Theoretical Physics and Mathematics, P.O. Box 19395-5746, Tehran,
Iran }}
\begin{document}
\maketitle
\begin{abstract}
We study the cosmological aspects of a noncommutative,
multidimensional universe where the matter source is assumed to be a
scalar field which does not commute with the internal scale factor.
We show that such noncommutativity
results in the internal dimensions being stabilized.\vspace{3mm}\noindent
\end{abstract}\vspace{5mm}

\section{Introduction}
Multidimensional theories of general relativity and cosmology are
an active area of research and have become even more so in the
past few decades. As a key reason for this interest one may point
to the common roots which such theories have with string theory
\cite{1} and its generalization, the M-theory \cite{2}. Also,
their success in describing the mass hierarchy problem \cite{3},
the cosmological constant problem \cite{4} and unification of
interactions have been notable in the recent past. However, as our
observed universe seems to have only four dimensions, at least at
energies  below the unification scale, multidimensional theories
face the challenging task of dealing with these internal
dimensions, that is, making them appropriately small and stable. A
myriad of explanations and solutions to these problems have been
suggested since the birth of such theories \cite{nima,5}.

One approach in this regard is that of the introduction of
noncommutativity into the theory. However, one should be careful to
distinguish noncommutativity in the geometry of space-time from that
between the fields.  Snyder did the first work on noncommutative
space-time \cite{snyder} which stirred a large amount of interest
\cite{connes,var,dog}. This interest has some roots in string and
M-theories \cite{string,m}. These noncommutative models have been
able to offer interesting results in dealing with problems such as
IR/UV mixing and non-locality \cite{6}, Lorentz violation \cite{7}
and new physics at very short distance scales \cite{8}. A different
approach is the introduction of noncommutativity between the fields
\cite{carmona}. Noncommutative cosmology \cite{garcia,babak} is an
example of such an approach and has benefited from it greatly in
that it can offer a view of the semiclassical approximation of
quantum gravity and may be used in tackling the cosmological
constant problem \cite{10}. Also, this kind of noncommutativity is
used in \cite{nima1} to address the stabilization of internal
dimensions and the cosmological constant problem.

In this paper, we have considered a multidimensional cosmology with
two scale factors for external and internal spaces in the presence
of a matter field. We show that if the matter field does not commute
with the scale factor associated with the internal dimensions, an
interesting result, namely stabilization of the internal dimensions
emerges. In our approach, in contrast to other approaches
\cite{nima,zhuk,zzz,sean}, the existence of a potential term  with
specific properties corresponding to the scalar field is
unnecessary. It will be shown that noncommutativity plays a crucial
role in the study of stabilization of the internal dimensions. It
should however be remarked that the model presented here is a toy
model in that in a more comprehensive setting, one should consider a
more general metric for which the brackets of all the components
with the scalar field should be calculated and taken into account.

\section{The Model}
Consider a cosmological model on a manifold defined as
\begin{equation}
M=R\times M_1 \times M_2 ,
\end{equation}
with a metric of the form
\begin{equation}\label{1}
g=-e^{2\gamma(\tau)}d\tau \otimes d\tau+e^{2u(\tau)}dx^i \otimes
dx^i+e^{2v(\tau)}dx^\alpha \otimes dx^\alpha ,
\end{equation}
where $i=1,2,3$ and $\alpha=1,2,...,d$ with $d$ being the
dimension of the internal space. Note that the above homogeneous
metric is also isotropic with different scale factors with respect
to the internal and ordinary spaces separately. These models are
natural multidimensional generalizations of the Friedmann as well
as Kasner universes \cite{1}. Here, we investigate a model with a
cosmological constant $\Lambda$ and a non-interacting homogenous
minimally coupled scalar field $\phi$ represented by the potential
$U(\phi)$. The action functional can be written as
\begin{equation}\label{2}
{\cal{S}}=\frac{1}{2\kappa^2}\int d^Dx \sqrt{g}\left\{
R[g]-2\Lambda\right\}+{\cal{S_\phi}}+{\cal{S}}_{YGH},
\end{equation}
where $R[g]$ is the scalar curvature of the metric (\ref{1}),
$\kappa^2$ is the $D$-dimensional gravitational constant, where
$D=d+3$ and ${\cal{S}}_{YGH}$ is the usual York-Gibbons-Hawking
boundary term.  The action of the matter field $\cal{S_\phi}$ is
\begin{equation}
{\cal{S_\phi}}=\int d^Dx
\sqrt{g}\left[-\frac{1}{2}g^{\mu\nu}\partial_\mu\phi\partial_\nu\phi-
U(\phi)\right].
\end{equation}
Assuming that all fields are functions of $\tau$ and using metric
(\ref{1}), action (\ref{2}) can be written as
\begin{equation}
{\cal{S}}=k\int d\tau L ,
\end{equation}
where $k$ is a constant and
\begin{equation}
L=-\frac{1}{2}e^{-\gamma+\gamma_0}\left[6\dot{u}^2+(d^2-d)\dot{v}^2+
6d\dot{u}\dot{v}-\kappa^2\dot{\phi}^2\right]
-e^{\gamma+\gamma_0}\left[\kappa^2U(\phi)+\Lambda\right].
\end{equation}
Here, a dot represents  differentiation with respect to $\tau$ and
$\gamma_0=3u+dv$. Choice of the harmonic time gauge as
$\gamma=\gamma_0$ \cite{11} results in
\begin{equation}
L=-\frac{1}{2}\left[6\dot{u}^2+(d^2-d)\dot{v}^2+6d\dot{u}\dot{v}-\kappa^2\dot{\phi}^2\right]
-e^{2(3u+dv)}\left[\kappa^2U(\phi)+\Lambda\right].
\end{equation}
For the above Lagrangian one may write the corresponding
Hamiltonian as
\begin{equation}\label{3}
H=\alpha p_u^2+\beta p_v^2 + \delta
p_up_v+\frac{1}{2\kappa^2}p_\phi^2+
e^{2(3u+dv)}(\kappa^2U(\phi)+\Lambda),
\end{equation}
where $p_u$, $p_v$ and $p_\phi$ are the momenta conjugate to $u$,
$v$, and $\phi$ respectively and $\alpha$, $\beta$, and $\delta$
are the following constants
\begin{eqnarray}
\alpha&=&\frac{d-1}{6(d+2)},\nonumber\\
\beta&=&\frac{1}{d(d+2)},\\
\delta&=&\frac{-1}{d+2}.\nonumber
\end{eqnarray}
Equations of motion correspond to Hamiltonian (\ref{3}) become
\begin{eqnarray}\label{4}
\dot{u}&=&\{u,H\}_P=2\alpha p_u+\delta p_v,\nonumber\\
\dot{p_u}&=&\{p_u,H\}_P=-6V,\nonumber\\
\dot{v}&=&\{v,H\}_P=2\beta p_v+\delta p_u,\nonumber\\
\dot{p_v}&=&\{p_v,H\}_P=-2dV,\\
\dot{\phi}&=&\{\phi,H\}_P=\frac{1}{\kappa^2}p_\phi,\nonumber\\
\dot{p_\phi}&=&\{p_\phi,H\}_P=-\kappa^2e^{2(3u+dv)}\frac{dU(\phi)}{d\phi},\nonumber
\end{eqnarray}
where $V=e^{2(3u+dv)}(\kappa^2U(\phi)+\Lambda)$ is the effective
potential. Now, using equations (\ref{4}), we obtain
\begin{eqnarray}\label{5}
\ddot{u}&=&-2(6\alpha +d\delta)V,\nonumber\\
\ddot{v}&=&-2(2d\beta +3\delta)V,\\
\ddot{\phi}&=&-e^{2(3u+dv)}\frac{dU(\phi)}{d\phi}.\nonumber
\end{eqnarray}
To address the stabilization problem one usually assumes the
existence of a potential $U(\phi)$ with a minimum at a definite
value. In this mechanism, in the neighborhood of the minimum,
stabilization of the extra dimensions for Ricci-flat \cite{zhuk}
and non-flat  \cite{nima,zzz,sean} internal spaces can be
achieved. This mechanism does not work in our model, the reason
being the special gauge chosen here. To be more specific, if we
use the mechanism used in \cite{zhuk}, the resulting solutions
obtained from equations (\ref{5}) may be written as
\begin{eqnarray}\label{11}
u&=&b_1\tau +b_2,\nonumber\\
v&=&b_3\tau +b_4,\\
\phi&=&b_5\tau +b_6,\nonumber
\end{eqnarray}
from which one can easily write the scale factors as
\begin{eqnarray}
R(\tau)=e^{2u(\tau)}=e^{2(b_1\tau +b_2)},\\
a(\tau)=e^{2v(\tau)}=e^{2(b_3\tau +b_4)}.
\end{eqnarray}
It is obvious that the scale factors diverge or converge according
to the sign of $b_1$ and $b_3$, and there would no stabilization
for the internal dimensions. In the next section we introduce the
notion of noncommutativity and show that in the harmonic time
gauge, it enables us to address the question of stabilization.
\section{Noncommutative solutions}
Let us now concentrate on the study of noncommutativity concepts
with Moyal product in phase space. The Moyal product may be traced
to an early intuition by Wigner \cite{wigner} which has been
developing over the past decades \cite{qqq}. Noncommutativity in
classical physics \cite{12} is described by the Moyal product law
between two arbitrary functions of position and momenta
\begin{equation}\label{35}
(f\star_\alpha
g)(x)=\mbox{exp}\left[\frac{1}{2}\alpha^{ab}\partial_a^{(1)}\partial_b^{(2)}
\right]f(x_1)g(x_2)|_{x_1=x_2=x},
\end{equation}
such that
\begin{eqnarray}\label{37}
\alpha_{ab}=
\left(%
\begin{array}{cc}
  \xi_{ij} & \delta_{ij}+\sigma_{ij} \\
 -\delta_{ij}-\sigma_{ij} & \zeta_{ij} \\
\end{array}%
\right)
\end{eqnarray}
where $N\times N$ matrices $\xi$ and $\zeta$ are assumed to be
antisymmetric. Also, $2N$ is dimension of  the classical phase
space $(x_i,p_i)$ and $i=1,2,...,N$. With this product law,
deformed Poisson brackets can be written as
\begin{equation}\label{37}
\{f,g\}_\alpha=f\star_\alpha g-g\star_\alpha f,
\end{equation}
where
\begin{eqnarray}\label{38}
\{x_i,x_j\}_\alpha&=&\xi_{ij},\nonumber \\
\{x_i,p_j\}_\alpha&=&\delta_{ij}+\sigma_{ij}, \\
\{p_i,p_j\}_\alpha&=&\zeta_{ij}. \nonumber
\end{eqnarray}
It is worth noting at this stage that in addition to
noncommutativity in $(x_i,x_j)$ we have also considered
noncommutativity in the corresponding momenta. This should be
interesting since its existence is in fact due essentially to the
existence of noncommutativity on the space sector \cite{qqq,12}
and it would somehow be natural to include it in our
considerations. Now, consider the following transformations in the
classical phase space $(x_i,p_i)$
\begin{eqnarray}\label{39}
\left\{
\begin{array}{ll}
{x_i'}=x_i-\frac{1}{2}\xi_{ij}p_j,\\
\\
{p_i'}=p_i+\frac{1}{2}\zeta_{ij}x_j.
\end{array} \right.
\end{eqnarray}
It can easily be checked that if $(x_i,p_i)$ obey the usual
Poisson algebra,  then
\begin{eqnarray}\label{40}
\{{x_i'},{x_j'}\}_{P}&=&\xi_{ij},\nonumber\\
\{{x_i'},{p_j'}\}_{P}&=&\delta_{ij}+\sigma_{ij},\\
\{{p_i'},{p_j'}\}_{P}&=&\zeta_{ij},\nonumber
\end{eqnarray}
where $\sigma_{ij}$ can be written as a combination of $\xi_{ij}$
and $\zeta_{ij}$. These commutation relations are the same as those
in (\ref{38}). Thus, for introducing noncommutativity, it is more
convenient to work with Poisson brackets (\ref{40}) than the
$\alpha$-star deformed Poisson brackets (\ref{38}). In the
noncommutative case we only assume that the scale factor associated
with the internal dimensions does not commute with the matter field
and keep other commutation relations unchanged as in the previous
section. Therefore, the relations that change are
\begin{eqnarray}
\left\{
\begin{array}{ll}
\{v',\phi'\}_P=\xi,\\
\\
\{p_v',p_\phi'\}_P=\zeta.
\end{array} \right.
\end{eqnarray}
The Hamiltonian $H'$ can be written as
\begin{equation}
H'=\alpha p_u'^2+\beta p_v'^2 + \delta
p_u'p_v'+\frac{1}{2\kappa^2}p_\phi'^2+
e^{2(3u'+dv')}\left[\kappa^2U(\phi')+\Lambda\right],
\end{equation}
where a prime denotes noncommutative variables. With the help of
transformations
\begin{equation}
\begin{array}{llll}
v'=v-\frac{\xi}{2}p_\phi\vspace{.5cm},\\
\phi'=\phi+\frac{\xi}{2}p_v\vspace{.5cm},\\
p_v'=p_v+\frac{\zeta}{2}\phi\vspace{.5cm},\\
p_\phi'=p_\phi-\frac{\zeta}{2}v\vspace{.5cm},\\
\end{array}
\end{equation}
we can write the Hamiltonian without the prime variables, that is,
those that satisfy the usual commutation relations
\begin{eqnarray}\label{h}
H&=&\alpha p_u^2+\beta p_v^2 + \delta
p_up_v+\frac{1}{2\kappa^2}p_\phi^2+\beta\zeta\phi
p_v+\frac{\delta\zeta}{2}\phi p_u-\frac{\zeta}{2\kappa^2}vp_\phi
+\frac{\beta\zeta^2}{4}\phi^2+\frac{\zeta^2}{8\kappa^2}v^2\nonumber\\
 &+&e^{2(3u+dv)}e^{-\xi
dp_\phi}\left[\kappa^2U(\phi+\frac{\xi}{2}p_v)+\Lambda\right].
\label{6}
\end{eqnarray}
The equations of motion can now be written easily with respect to
Hamiltonian (\ref{6})
\begin{equation}
\begin{array}{llllll}
\hspace{1.5cm}\dot{u}=\{u,H\}_P=2\alpha p_u+\delta p_v+\frac{\delta\zeta}{2}\phi,\\
\hspace{1.5cm}\dot{v}=\{v,H\}_P=2\beta p_v+\delta p_u+\beta\zeta\phi+e^{2(3u+dv)}
e^{-\xi dp_\phi}\kappa^2\frac{\partial U(\phi+\frac{\xi}{2}p_v)}{\partial p_v},\\
\hspace{1.5cm}\dot{\phi}=\{\phi,H\}_P=\frac{1}{\kappa^2}p_{\phi}-\frac{\zeta}{2\kappa^2}v-\xi
de^{2(3u+dv)}e^{-\xi dp_\phi}\left[\kappa^2U(\phi+\frac{\xi}{2}p_v)+\Lambda\right],\\
\hspace{1.5cm}\dot{p_u}=\{p_u,H\}_P=-6e^{2(3u+dv)}e^{-\xi dp_\phi}
\left[\kappa^2U(\phi+\frac{\xi}{2}p_v)+\Lambda\right],\\
\hspace{1.5cm}\dot{p_v}=\{p_v,H\}_P=\frac{\zeta}{2\kappa^2}p_{\phi}-\frac{\zeta^2}{4\kappa^2}v-
2de^{2(3u+dv)}e^{-\xi dp_\phi}\left[\kappa^2U(\phi+\frac{\xi}{2}p_v)+\Lambda\right],\\
\hspace{1.5cm}\dot{p_\phi}=\{p_\phi,H\}_P=-\beta\zeta
p_v-\frac{\delta\zeta}{2}p_u-\frac{\beta\zeta^2}{2}\phi-\kappa^2e^{2(3u+dv)}e^{-\xi
dp_\phi}\frac{\partial U(\phi+\frac{\xi}{2}p_v)}{\partial \phi}.\\
\end{array}
\end{equation}
To proceed any further, we expand the potential about its minimum
taken at $-\Lambda/\kappa^2$, obtaining
\begin{eqnarray}\label{9}
 \ddot{u}&=&\delta\zeta\dot{\phi},\nonumber\\
 \ddot{v}&=&2\beta\zeta\dot{\phi},\\
 \ddot{\phi}&=&-\frac{\zeta}{\kappa^2}\dot{v},\nonumber
\end{eqnarray}
whose solutions are
\begin{eqnarray}\label{10}
\phi&=&\frac{c_1}{\omega}\sin(\omega \tau)-
\frac{c_2}{\omega}\cos(\omega \tau)+c_3,\nonumber\\
u&=&-\frac{c_1}{\omega^2}\delta\zeta \cos(\omega\tau)-
\frac{c_2}{\omega^2}\delta\zeta \sin(\omega\tau)+\frac{1}{2}k\tau+c_4,\\
v&=&-\frac{c_1}{\omega^2}2\beta\zeta \cos(\omega\tau)-
\frac{c_2}{\omega^2}2\beta\zeta \sin(\omega\tau)+c_5,\nonumber
\end{eqnarray}
where $c_1, c_2, c_3, c_4, c_5$ and $k$ are integration constants
and
\begin{equation}
\omega^2=\frac{2\beta\zeta^2}{\kappa^2}.\\
\end{equation}
Comparison between these solutions and those presented in
(\ref{11}) reveals the existence of  oscillating terms and a
linear term in $\tau$ which only appears in  the scale factor $u$.
It is worth noting that even if there was no effective potential
$V$ present, the above discussions and results would still hold.
\begin{figure}
\centerline{\begin{tabular}{c}
 \epsfig{figure=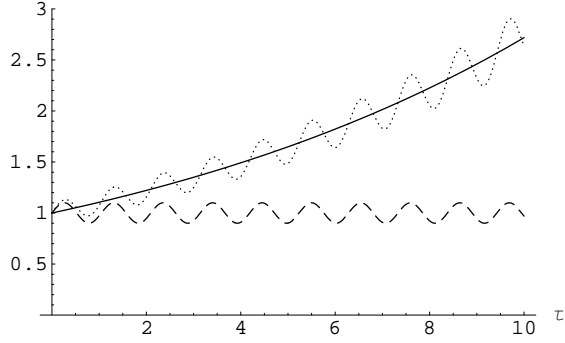,width=8cm}
 \end{tabular}  }
\caption{\footnotesize Commutative scale factors, solid line,
ordinary noncommutative scale factor, dotted line and noncommutative
scale factor associated with the internal dimensions, dashed line.
Note that the solid line represents both scale factors in the
commutative case.} \label{fig1}
\end{figure}

Let us now study the behavior of these solutions. For simplicity
we take  $c_1=0$ so that the scale factor of internal dimensions
can be written as
\begin{equation}\label{21}
a(\tau)=e^{2v(\tau)}=e^{2c_5}\exp\left[-\frac{4c_2}{\omega^2}\beta\zeta
\sin(\omega\tau)\right].\\
\end{equation}
This means that the scale factor of the internal dimensions is
stabilized around $e^{2c_5}$. On the other hand, the other scale
factor reads
\begin{equation}\label{22}
R(\tau)=e^{2u(\tau)}=
e^{2c_4}\exp\left[-\frac{2c_2}{\omega^2}\delta\zeta
\sin(\omega\tau)\right]e^{k\tau}.
\end{equation}
The interpretation of equation (\ref{22}) now depends on the sign of
$k$. To have an understanding of this solution let us look at the
Hubble constant $\textsf{H}$ for the scale factor of ordinary
universe in noncommutative case where it can be written as
\begin{equation}\label{23}
\textsf{H}=\frac{\dot{R}}{R}=k-2c_2\frac{ \delta
\zeta}{\omega}\cos(\omega\tau).
\end{equation}
The average of the Hubble constant (\ref{23}) during its period is
equal to $k$. Since observations show that the Hubble constant is
positive, a positive sign for $k$ in equation (\ref{22}) would also
seem appropriate and natural. Having taken a positive value for $k$,
one would immediately see that equation (\ref{22}) would result in a
large value for the scale factor of the universe consistent with
present observations. This equation also shows that the usual scale
factor has an oscillatory behavior and that the amplitude of
oscillations grows exponentially. The behavior of the scale factors
can be seen in figure 1. It should be noted that in our
noncommutative model, there is no requirement for the existence of a
potential with a minimum, necessary to describe stabilization in
other models \cite{nima,zhuk,zzz,sean}. Therefore, stabilization
seems to have a somewhat deeper roots in noncommutativity than in
the potential. In other words, the role of such a potential is
played by the term
$\frac{\beta\zeta^2}{4}\phi^2+\frac{\zeta^2}{8\kappa^2}v^2$,
appearing spontaneously in (\ref{h}) and is due to the introduction
of noncommutativity. This term has a minimum at $\phi_0=0$ and
$v_0=0$ where upon the stabilization conditions discussed in
\cite{nima,zhuk,zzz,sean} are satisfied.

Recent observations seem to suggest that the universe has two
distinct decelerated and accelerated phases (see the first
reference in \cite{dark}). To account for this behavior, models
based on dark energy \cite{dark} and modified gravity \cite{mod}
have been suggested. To address such questions in our simple
model, let us calculate the acceleration parameter $q$
\begin{eqnarray}
q=\frac{\ddot{R}R}{\dot{R}^2}=1+2
c_2\beta\zeta\frac{\sin(\omega\tau)}{\textsf{H}}. \label{233}
\end{eqnarray}
It should now be clear that the desired behavior, that is positive
or negative values for $q$ corresponding to accelerated or
decelerated phases respectively, can be obtained from the above
relation by tuning the parameters. Since the existence of a time
dependent term in equation (\ref{233}) is a direct consequence of
noncommutativity, such phases could be interpreted as the direct
result of  this effect. Also,  equation (\ref{233}) suggests a
periodic phase transition, as can be seen in figure 1.

\section{Conclusions}
In this paper, we have investigated a multidimensional
cosmological model with two different scale factors in the
presence of a massless scalar field. In addition, we have assumed
that the scale factor of the internal space does not commute with
that of the matter field. The motivation for this assumption is
that in different approaches to multidimensional theories, the
internal space scale factors may be considered as matter fields
\cite{13,zzz}. So from the viewpoint of an observer in ordinary
universe, there is no distinction between matter fields and scale
factors of the internal space. Hence noncommutativity between
these fields seems to be a natural choice. With these assumptions,
it can be seen from equations (\ref{21},\ref{22}) that at early
times or small $\tau$, the two scale factors were in the
oscillating regime and can be treated as having sizes of the same
order of magnitude. However, with time increasing, because of the
exponential term in (\ref{22}), the ordinary space scale factor
begins to move away in size from the internal dimensions scale
factor. So, for large $\tau$, that is for the present epoch, the
scale factor of the observed universe has grown exponentially in
contrast to the scale factor of the internal dimensions which are
and have been oscillatory. One may therefore speaks of the
stabilization of the internal dimensions in the present epoch
around a small constant radius. It should be interesting to note
that the various models suggested for stabilization
\cite{nima,zhuk,zzz,sean} do not work in the harmonic time gauge
whereas the notion of noncommutativity proposed in this paper
points in the direction of offering a solution. In addition, it
suggests a way to describe the two different phases associated
with our universe mentioned above.
\vspace{5mm}\noindent\\
{\bf Acknowledgement}\vspace{1mm}\noindent\\
We would like to thank B. Vakili for a careful reading of the
manuscript.

\end{document}